\documentclass{mn2e}
\usepackage{graphicx}

\begin{document}

\title{On the relationship between the $\delta$ Scuti and $\gamma$
Doradus pulsators}
\author[G. Handler \& R. R. Shobbrook]
    {G. Handler$^{1}$\thanks{E-mail: gerald@saao.ac.za} and
	R. R. Shobbrook$^{2,3}$ 
	\and \\
$^{1}$ South African Astronomical Observatory, P.O. Box 9, Observatory 7935,
South Africa\\
$^{2}$ P. O. Box 518, Coonabarabran, N.S.W 2357, Australia\\
$^{3}$ Research School of Astronomy and Astrophysics,
Australian National University, Weston Creek P.O., ACT 2611, Australia}

\date{Accepted 2001 nnnn nn.
   Received 2001 nnnn nn;
   in original form 2001 nnnn nn}

\maketitle

\begin{abstract} 

We searched for $\delta$~Scuti-type pulsations amongst known and
candidate $\gamma$ Doradus stars. The motivations for such a project come
from the need to understand the relationship of these two classes of
pulsator better, from the present poor knowledge of the hot border of the
$\gamma$ Doradus phenomenon, and from the exciting prospects for
asteroseismology should stars be found which have both types of pulsation
excited.

We acquired 270 h of observations and monitored a total of 26 stars. One
target, HD 209295, turned out to be a member of both classes of pulsating
star, but this object is peculiar in the sense that it is a close binary.
We classify six of our targets as new {\it bona fide} $\gamma$ Doradus
stars, whereas nine more are good $\gamma$ Doradus candidates, and three
turned out to be ellipsoidal variables. One of our program stars was found
to be a $\delta$~Scuti star, with no additional $\gamma$ Doradus
variations. Furthermore, one star was already known to be a {\it bona
fide} $\gamma$ Doradus star, and we could not find an unambiguous
explanation for the variability of five more stars.

The analysis of our data together with improved knowledge of stars from
the literature enabled us to revise the blue border of the $\gamma$
Doradus phenomenon towards cooler temperatures. This new blue edge is much
better defined than the previous one and extends from a temperature of
about 7550 K on the ZAMS to 7400 K one magnitude above it.

Five {\it bona fide} $\gamma$ Doradus stars we observed are located inside
the $\delta$ Scuti instability strip, but none of them exhibited
observable $\delta$ Scuti pulsations. We therefore suggest that $\gamma$
Doradus stars are less likely to be $\delta$~Scuti pulsators compared to
other normal stars in the same region of the lower instability strip. In
addition, we show that there is a clear separation between the pulsation
constants $Q$ of $\delta$ Scuti and $\gamma$ Doradus stars. The $\gamma$
Doradus stars known to date all have $Q>0.23$~d.

\end{abstract}


\section{Introduction}

The relationship between the $\gamma$ Doradus and $\delta$ Scuti stars is
not yet clear. Although the two classes of pulsator share a similar
parameter space in the HR diagram and even partly overlap (Handler 1999),
the $\gamma$ Doradus stars are high radial-order gravity (g)-mode
pulsators (Kaye et al. 1999a), whereas the $\delta$ Scuti stars are
believed to be mostly low radial-order pressure (p) mode pulsators (e.g.
Breger 2000).

The driving mechanism of the two classes should be different because of
the different types of modes excited. Hence, there should be different
thermal time scales in the driving regions. Indeed, $\delta$ Scuti
pulsations are known to be driven by the $\kappa$-mechanism (Chevalier
1971), whereas the only presently feasible driving mechanism for the
$\gamma$ Dor stars is similar to convective blocking (Guzik et al. 2000,
2002). The latter mechanism is also expected to weaken or even exclude the
driving of $\delta$ Scuti-type pulsations.

Several open questions still remain. Is the presence of $\delta$ Scuti and
$\gamma$ Doradus-type pulsations mutually exclusive? Is there an overlap
or a separation between these pulsators? Are there stars which show both
types of oscillation? What determines the type of mode that a particular
star pulsates in? To tackle those questions, we decided to observe a large
number of $\gamma$ Doradus candidates located within the $\delta$ Scuti
instability strip to search for short-period pulsations.

In addition, this allows us to check whether these $\gamma$ Doradus
candidates are indeed pulsators. This is important as most of these
candidates were identified from {\it Hipparcos} photometry (ESA 1997), whose
quality is poor compared to ground-based work and which contains no
time-series colour information. Furthermore, there is usually strong
aliasing in amplitude spectra of {\it Hipparcos} photometry caused by a
pseudo-sampling frequency near 12 c/d (see Eyer \& Grenon 1998), which
often leads to difficulties in period determinations. For instance it is
often not possible to distinguish a $\delta$ Scuti star with a 2-hour
period from a $\gamma$ Doradus candidate from {\it Hipparcos} photometry only.
We conclude that these technical facts make a good assessment of the
presence of $\gamma$ Doradus pulsations from {\it Hipparcos} data alone
difficult: $\gamma$ Doradus stars can easily be confused with spotted
stars, ellipsoidal variables and $\delta$ Scuti stars (see Kaye et al.
1999a and Handler 1999 for more extensive discussions).

Our ground-based observations introduced above are probably the most
effective way to assess the relationship between the $\gamma$ Doradus and
$\delta$ Scuti stars: about one third of all non-Am and non-Ap stars in
the lower instability strip are indeed $\delta$ Scuti stars (Breger 2000
and references therein) and Am and Ap stars are rare (or even absent)
amongst the $\gamma$ Doradus stars (Handler 1999). Consequently, a
statistical assessment of the incidence of $\delta$ Scuti pulsations in
$\gamma$ Doradus stars becomes possible. We note that Breger \&
Beichbuchner (1996) already looked for possible $\gamma$ Doradus pulsators
among $\delta$ Scuti stars in the literature, but their results were
largely inconclusive because of observational selection effects, the small
number of $\gamma$ Doradus stars then known and because the incidence of
observable $\gamma$ Doradus pulsations among stars with basic parameters
in the $\gamma$ Doradus star domain is as yet unknown.

The results to be expected from our search are manifold.
\begin{itemize}

\item The number of known {\it bona fide} $\gamma$ Doradus stars is still
small ($< 20$). Discovering more of these stars will aid the understanding
of the class as a whole.
\item If both types of pulsation could be identified in the same star, 
this would be exciting news for asteroseismology: the $\delta$ Scuti 
pulsations in such hypothesized stars could be used to determine (at 
least) accurate basic parameters, which can then aid in identifying the
$\gamma$ Doradus modes and in probing the deep stellar interior. It would
also mean that in stellar models in which $\gamma$ Doradus-type modes are
driven, $\delta$ Scuti oscillations must not be damped.
\item If no such ``hybrid'' stars are found, the reason for their absence
can be examined, and some observational discriminants may be found. Of
course, the driving mechanism for the $\gamma$ Doradus stars will also
need to explain such a result.
\item An improved assessment of the hotter candidate $\gamma$ Doradus 
stars may also enable us to locate the currently poorly defined blue 
edge of the $\gamma$ Doradus region in the HR diagram (Handler 1999) 
better; pulsational models must explain it. The location of the red
edge of the $\delta$ Scuti instability strip can also be better examined.
\end{itemize}

Hence, whatever the implications of the present survey, the driving
mechanism for the $\gamma$ Dor pulsations will be observationally much
better constrained.

\section{Observations}

As mentioned in the Introduction, we selected known candidate $\gamma$
Doradus stars located within the overlap region of the lower instability
strip (adopted from Breger 1979) and the domain of the $\gamma$ Doradus
stars in the HR diagram (Handler 1999). All stars observable from
intermediate Southern geographical latitudes were included in our survey,
and we chose two comparison stars for each. In addition, we included
further $\gamma$ Doradus candidates (regardless of their position in the
HR diagram) which were located within 8 degrees of the main targets into
the observing sequence whenever practical to examine the physical nature
of these stars as well. Finally, we also decided to monitor a few more
candidates which seemed to be of special astrophysical interest.

Our measurements were acquired as differential photoelectric photometry at
the 0.5-m and 0.75-m telescopes at the Sutherland station of the South
African Astronomical Observatory (SAAO) and at the 0.6-m telescope (see
Shobbrook 2000 for a description) at Siding Spring Observatory (SSO) from
October 1999 to October 2001. We used the Johnson B and V filters with a
total integration time of about one minute in each filter as a compromise
between good time resolution and optimum colour information, but we
sometimes added the Cousins I$_{\rm c}$ if deemed necessary (possible
reasons will be listed below). A few runs obtained on the SAAO 0.5-m
telescope were taken as high-speed photometric observations through the
BVI$_{\rm c}$ filters, with comparison star observations every 30 - 45
minutes, which also results in some long-term photometric stability
(Breger \& Handler 1993). For all runs, apertures of 30 -- 45 arcseconds
on the sky were chosen. Sky measurements were taken depending on the
brightness and proximity of the Moon.

Observing sequences were chosen for both good coverage of possible
short-period $\delta$ Scuti pulsations and for best long-term stability.
We adopted the sequences C1-C2-V-C1-C2-V... (the Cs denote the comparison
stars and the Vs the variables) for SAAO observations and C1-V-C2-V-C1...
for SSO measurements (where it was more difficult to move the telescope
from one star to the next) in case one programme star was in the group.
This resulted in one programme star measurement about every 7 minutes. If
we had two variables in a group, the sequence C1-V1-C2-V2-C1-V1... was
chosen, yielding a variable star measurement about every 15 minutes. These
duty cycles apply to BV observations.

We attempted to observe each target star for at least two half nights to
check them for $\delta$ Scuti-type variability and in order not to be
susceptible to beating phenomena (negative interference of multiple
pulsation modes). With this strategy, long-term variability is often
detected in one night, and the second run will fulfil the same aim whilst
making it possible to check for light variations from night to night as
well. The data were reduced as soon as possible after they had been
obtained in order to judge their quality and to be able to make decisions
about future observing strategies and the scientific content of our data.
This would for instance result in a change of the observing sequence or in
the inclusion of the I$_{\rm c}$ filter or in a decision of whether to
follow the star up or not.

Data reduction was performed in a standard way: correction for coincidence
losses, sky background and extinction was followed by calculating
differential magnitudes between the comparison stars, and if these were
judged to be constant, construction of the differential target star light
curve. Finally, the time base of our observations was converted to
Heliocentric Julian Date (HJD). We summarize our observations in Table 1;
the data are available in electronic form at {\tt
www.saao.ac.za/$\sim$gerald/delgamscudor}.

\begin{table}
\caption[]{Journal of the differential photometric observations. We 
normally used the Johnson BV filters, but runs marked with one 
asterisk also utilised the I$_{\rm c}$ filter; those with two
asterisks were high-speed photometry BVI$_{\rm c}$ with occasional
comparison star measurements.}
\begin{flushleft}
\begin{tabular}{llcc}
\hline
Target star(s) & Site & Run start & Run length \\
 & & JD - 2450000 & h \\
\hline
HD 10167 & SAAO & 1457.40 & 4.1\\
 & SSO & 2111.19 & 3.1\\
 & SSO & 2112.16 & 1.6\\
 & SSO & 2131.16 & 2.6\\
HD 12901 & SAAO & 1460.38 & 4.1\\
 & SAAO & 2154.48 & 2.0$^{\ast\ast}$\\
 & SAAO & 2155.47 & 4.5$^{\ast\ast}$\\
HD 14147 & SAAO & 1463.39 & 3.0\\
HD 27093 & SAAO & 1461.25 & 3.7\\
 & SAAO & 1466.48 & 2.2\\
HD 40745 and & SSO & 1549.96 & 5.4 \\
HD 41448 & SSO & 1554.96 & 1.9 \\
 & SSO & 1555.95 & 5.1 \\
HD 65526 & SAAO & 1576.30 & 3.5 \\
 & SAAO & 1581.30 & 3.7 \\
HD 81421 & SAAO & 1577.32 & 2.3\\
 & SAAO & 1578.31 & 6.2\\
 & SSO & 1578.95 & 7.0\\
 & SAAO & 1579.30 & 7.3\\
 & SAAO & 1581.57 & 1.0\\
 & SAAO & 1582.54 & 1.7\\
 & SSO & 1588.93 & 4.0\\
 & SSO & 1589.96 & 5.6\\
 & SSO & 1590.97 & 4.7\\
HD 85693 and & SAAO & 1580.30 & 3.8\\
HD 86371 & SAAO & 1582.33 & 5.0\\
HD 110606 and & SAAO & 1576.46 & 2.9\\
HD 113357 & SAAO & 1580.45 & 0.4\\
 & SAAO & 1581.46 & 2.7\\
 & SSO & 1621.12 & 4.1\\
 & SSO & 1630.00 & 2.6\\
 & SSO & 1631.96 & 2.9\\
HD 139095 & SAAO & 2045.29 & 1.6\\
 & SSO & 2102.85 & 6.1\\
 & SSO & 2110.86 & 6.2\\
 & SSO & 2111.86 & 5.1\\
HD 167858 & SAAO & 2045.44 & 5.4\\
 & SAAO & 2051.54 & 3.1\\
BD+8 3658 & SSO & 2105.01 & 1.6\\
 & SSO & 2130.90 & 4.3\\
 & SSO & 2134.94 & 3.2\\
 & SSO & 2140.88 & 2.6$^{\ast}$\\
HD 173794 & SSO & 2131.87 & 8.3\\
 & SSO & 2133.12 & 1.0\\
 & SSO & 2135.09 & 2.6\\
 & SSO & 2135.87 & 7.7$^{\ast}$\\
 & SSO & 2140.99 & 3.0$^{\ast}$\\
 & SSO & 2144.88 & 7.1$^{\ast}$\\
HD 181998 & SSO & 2146.87 & 1.4\\
 & SAAO & 2154.33 & 3.4$^{\ast\ast}$\\
 & SAAO & 2155.24 & 5.6$^{\ast\ast}$\\
HD 188032 and & SAAO & 1458.26 & 2.5\\
HD 189631 & SSO & 2107.10 & 1.9\\
 & SSO & 2113.03 & 5.2\\
 & SAAO & 2192.23 & 2.9$^{\ast}$\\
 & SAAO & 2193.30 & 0.6$^{\ast}$\\
\hline
\end{tabular}
\end{flushleft}
\end{table}

\setcounter{table}{0}

\begin{table}
\caption[]{(Contintued)}
\begin{flushleft}
\begin{tabular}{llcc}
\hline
Target star(s) & Site & Run start & Run length \\
 & & JD - 2450000 & h \\ 
\hline
HD 198528 and & SAAO & 1461.25 & 1.1\\
HD 201985 & SAAO & 1462.25 & 3.6 \\
 & SAAO & 1466.26 & 3.6 \\
 & SSO & 2119.10 & 4.1\\
 & SSO & 2119.95 & 6.3\\
HD 207223 & SAAO & 1463.25 & 3.2\\
HD 207651 & SAAO & 1463.25 & 3.2\\
 & SAAO & 2191.26 & 3.0$^{\ast\ast}$\\
HD 209295 & SAAO & 1464.33 & 4.2\\
 & SAAO & 1465.25 & 6.5\\
 & SSO & 1467.03 & 2.3\\
 & SSO & 1468.91 & 6.4\\
HD 211699 & SAAO & 1468.25 & 1.1\\ 
 & SAAO & 1469.26 & 4.5\\ 
 & SAAO & 2176.26 & 2.1$^{\ast}$\\
 & SAAO & 2190.26 & 4.2$^{\ast\ast}$\\
HD 221866 & SAAO & 1466.41 & 1.6\\
 & SAAO & 1467.28 & 4.6\\
\hline
Total & & & 269.5\\
\hline
\end{tabular}
\end{flushleft}
\end{table}

\section{Analysis}

\subsection{Results on the individual stars}

Before proceeding to the astrophysical implications of our survey as a
whole, the individual stars also need to be discussed, as a wide variety
of behaviour was detected. We also need to make clear which criteria we
use to distinguish between the different types of variable to be found in
the region of the HR diagram under consideration.

We classify a star as a $\delta$ Scuti variable if it shows a variability
time scale which leads to pulsation constants $Q$ of 0.04 days or less.
The pulsation constant $Q$ is calculated as
\begin{equation}
\log Q_{\rm i} = C + 0.5\log g + 0.1M_{\rm bol} + \log T_{\rm eff} +
\log P_{\rm i},
\end{equation}
adopting $T_{\rm eff,\odot}$ = 5780 K, log $g_{\odot}$ = 4.44,
$M_{\rm bol,\odot}$ = 4.75 (Allen 1976) and thus $C = -6.456$. The units
of the quantities in the right-hand side of this equation are dex,
magnitudes, Kelvin and days, respectively. Our limit on $Q$ follows from
the fundamental radial mode of a $\delta$ Scuti star having
$Q=0.033$\,d, and we allow for some 18\% error in its determination from
the uncertainties of the physical parameters derived for the star (see
Breger 1989 for a discussion). These parameters can be determined from
calibrations of Str\"omgren photometry (Crawford 1975, 1979) and model
atmosphere predictions (Kurucz 1991). Bolometric corrections were taken
from Drilling \& Landolt (2000).

There are also possibilities to separate $\gamma$ Doradus stars,
ellipsoidal variables and rotationally modulated chemically peculiar
objects. The latter two types of variable will only have one or two
dominant periods in a frequency analysis based on Fourier spectra and
sine-wave fitting (which are our main tools for the analysis), and if
there are two periods, they will be harmonically related. In that respect,
residualgram analysis (Martinez \& Koen 1994) can become powerful. This
method performs a least-squares fit of a sine wave with $M$ harmonics to
the measured time series and evaluates the residuals at each trial
frequency, where $M$ can be chosen. For rotationally modulated light
curves or those of ellipsoidal variables, $M=2$ is usually the best
choice, and is used throughout this paper.

Returning to the discrimination between the different types of slow
variables, we note that multiperiodic $\gamma$ Doradus stars can be easily
identified if the different periods of variability are not harmonically
related. However, if only a single period can be determined from the
measurements, unravelling the cause of the variability of the star under
consideration becomes more difficult.

A further diagnostic can be obtained from relative amplitudes and phases
of the measured signals in different photometric filters. Ellipsoidal
variables or eclipsing binaries of the W UMa-type show little colour
modulation (B/V amplitude ratios less than 1.05) as their light variations
are dominated by geometrical effects. The latter two types of variable can
be distinguished by their light-curve shape; those of the W UMa-type have
typical flat maxima and sharp minima. The light curves of rotationally
modulated Ap stars, on the other hand, often show quite large colour
variations and in case of double-wave light modulations, phase shifts
between the different filters can become quite large (e.g. see Kurtz et
al. 1996). Colour amplitude ratios for $\gamma$ Doradus pulsations will be
quite similar to those of $\delta$ Scuti stars. For instance, typical B/V
amplitude ratios for $\gamma$ Doradus stars pulsating with photometrically
detectable modes (spherical degree $\ell=1$ or 2) would, according to
model calculations (e.g. Garrido 2000), be between 1.2 and 1.35. The
latter range is also expected for radial $\delta$ Scuti pulsation. Phase
shifts between different filters are not a good indicator of pulsations in
the present case, as our data sets are generally too small to obtain
significant phase shifts.

We caution, however, that amplitude ratios could be misleading, in
particular in case of insufficient phase coverage of the orbital period of
an ellipsoidal variable or of the rotation period of an Ap star. Both
types of variable can then show colour amplitude ratios similar to those
of pulsating stars. Consequently, this diagnostic alone is not sufficient
for a clear distinction; all the information available on the stars need
to be combined carefully to arrive at a safe classification.

\subsubsection{HD 10167} The variability of this star was discovered by
Eyer \& Aerts (2000) in Geneva photometry. These authors could however not
pinpoint the physical nature of the star. Our measurements show very
little variation during the individual nights and night-to-night
variations of a few hundreths of a magnitude. The $B/V$ colour amplitude
ratio (1.24 $\pm$ 0.12) suggests pulsation as the cause for these
variations. No $\delta$ Scuti-type variability is detected within a limit
of 0.9 mmag in the V-filter amplitude spectrum. We classify HD 10167 as a
$\gamma$ Doradus candidate.

\subsubsection{HD 12901} Handler (1999) suggested that this star could be
a $\gamma$ Doradus variable, but aliasing in the {\it Hipparcos} photometry also
left the possibility that it could be a $\delta$ Scuti star. Eyer \& Aerts
(2000) clarified the situation by means of Geneva photometry and suggested
that HD 12901 is a $\gamma$ Doradus star. Our observations confirm their
conclusion; no $\delta$ Scuti pulsations are detected (within 1.2 mmag)
and the $B/V$ and $V/I_{\rm c}$ colour amplitude ratios (1.29 $\pm$ 0.02
and 1.9 $\pm$ 0.1, respectively) for the dominant period of the slow
variations (0.8227 d, Eyer \& Aerts 2000) in our data confirm that they
are caused by pulsation of the star. Supported by the results of Eyer \&
Aerts (2000), we classify HD 12901 as a {\it bona fide} $\gamma$ Doradus
star.

\subsubsection{HD 14147} This star was classified as a $\delta$ Scuti star
with a period of 6.48 h by the {\it Hipparcos} group (ESA 1997). However,
a re-analysis of these data reveals that HD 14147 is in fact an
ellipsoidal variable with an orbital period of 12.95 h (Fig.~1). One can
suspect low-amplitude $\delta$ Scuti variations with a period of about 1.5
h ($Q=0.04$\,d) being superposed on our single night of observation and
the {\it Hipparcos} light curve.

\begin{figure}
\includegraphics[width=99mm,viewport=-10 00 335 345]{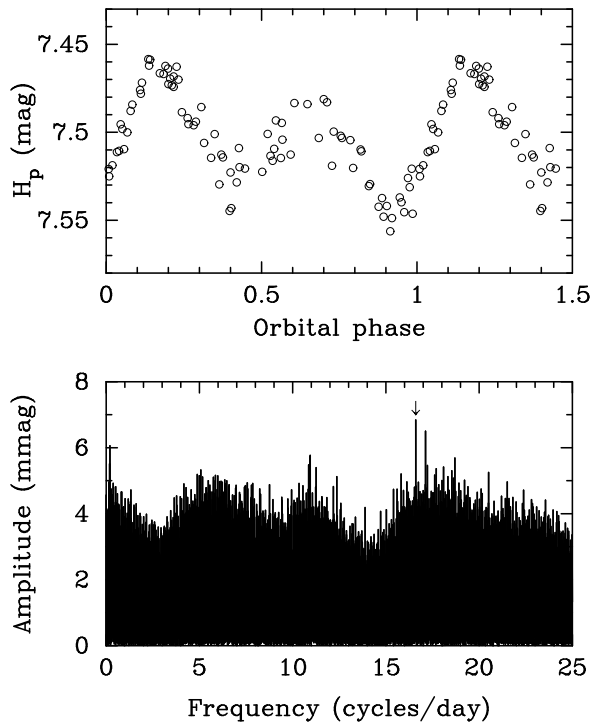}
\caption[]{Upper panel: phase diagram of the {\it Hipparcos} photometry of HD
14147 folded with a frequency of 1.8528 c/d. Ellipsoidal variability of
this star is strongly implied. Lower panel: residual amplitude spectrum of
this star's {\it Hipparcos} photometry after removing the orbital modulation. A
possible signal near 16.6 c/d can be suspected.}
\end{figure}

\subsubsection{HD 27093} From the time-resolved {\it Hipparcos} photometry
alone it was not clear whether this star is a $\delta$ Scuti or a $\gamma$
Doradus star (Handler 1999). However, the more rapid time sampling of our
ground-based observations clearly shows that HD 27093 is a $\delta$ Scuti
star; a combined analysis of both data sets results in a dominant
frequency of 14.51837 $\pm$ 0.00006 cycles/day, corresponding to
$Q=0.018$\d. Semi-amplitudes of 18 $\pm$ 2 mmag in the {\it Hipparcos}
$H_{\rm p}$ band, 15 $\pm$ 1 mmag in the B filter and with 11.8 $\pm$ 0.6
mmag in V could be determined (error estimates were derived from the
formulae of Montgomery \& O'Donoghue 1999). Some evidence for more
$\delta$ Scuti periods is seen, but no long-term variations are detected
within a limit of 2 mmag.

\subsubsection{HD 40745 and HD 41448} These two stars could be observed in
one group due to their proximity in the sky. Both are slow variables, with
B/V colour amplitude ratios (1.25 $\pm$ 0.21 for HD 40745 and 1.31 $\pm$
0.12 for HD 41448) indicative of pulsation. These amplitude ratios were
derived by fitting their dominant {\it Hipparcos} frequency to our data, which
are consistent with these frequencies. In accordance with published
frequency analyses of the {\it Hipparcos} data of the stars, we also find
evidence for multiperiodicity for the slow variations of both. On the
other hand, no $\delta$ Scuti-type variability is detected within a limit
of 3.0 mmag (HD 40745) and 2.7 mmag (HD 41448), respectively. Because of
the rather large errors on the colour amplitude ratios, we can only 
confirm both stars as prime $\gamma$ Doradus candidates.

\subsubsection{HD 65526} For this prime $\gamma$ Doradus candidate
(Handler 1999), we obtained two nights of observation which were of
excellent quality. The B/V colour amplitude ratio in our data (1.19 $\pm$
0.04) implies pulsation as the cause of the slow multiperiodic light
variations of HD 65526. No $\delta$ Scuti variations are detected at a
limit of 0.5 mmag in the amplitude spectrum. Thus HD 65526 is reclassified
as a {\it bona fide} $\gamma$ Doradus star.

\subsubsection{HD 81421} This is one of the most difficult cases we
encountered in our survey; we must comment on this star in more detail.
The star is classified as a periodic {\it Hipparcos} variable (ESA 1997),
and its period is 11.75 h (see also Fig.~2 (a)). This means that for
consecutive nights of ground-based measurements practically the same phase
of its light curve is always observed. For this reason, we coordinated the
observations of this star from both sites to overlap and we attempted to
cover a time base line longer than usual.

Regrettably, these attempts were somewhat unlucky as our combined
observations always covered the same branch of the light curve with the
exception of one night where we could observe it from both sites, albeit
with a 1.6-h gap in between. As we had no overlap between SSO and SAAO, we
transformed the instrumental magnitude differences to the standard system
to homogenize the data as much as possible. This procedure pointed us to a
zeropoint problem just in the single SSO night taken in between two nights
of SAAO data, which has to be kept in mind as well.

\begin{figure}
\includegraphics[width=102mm,viewport=5 00 330 619]{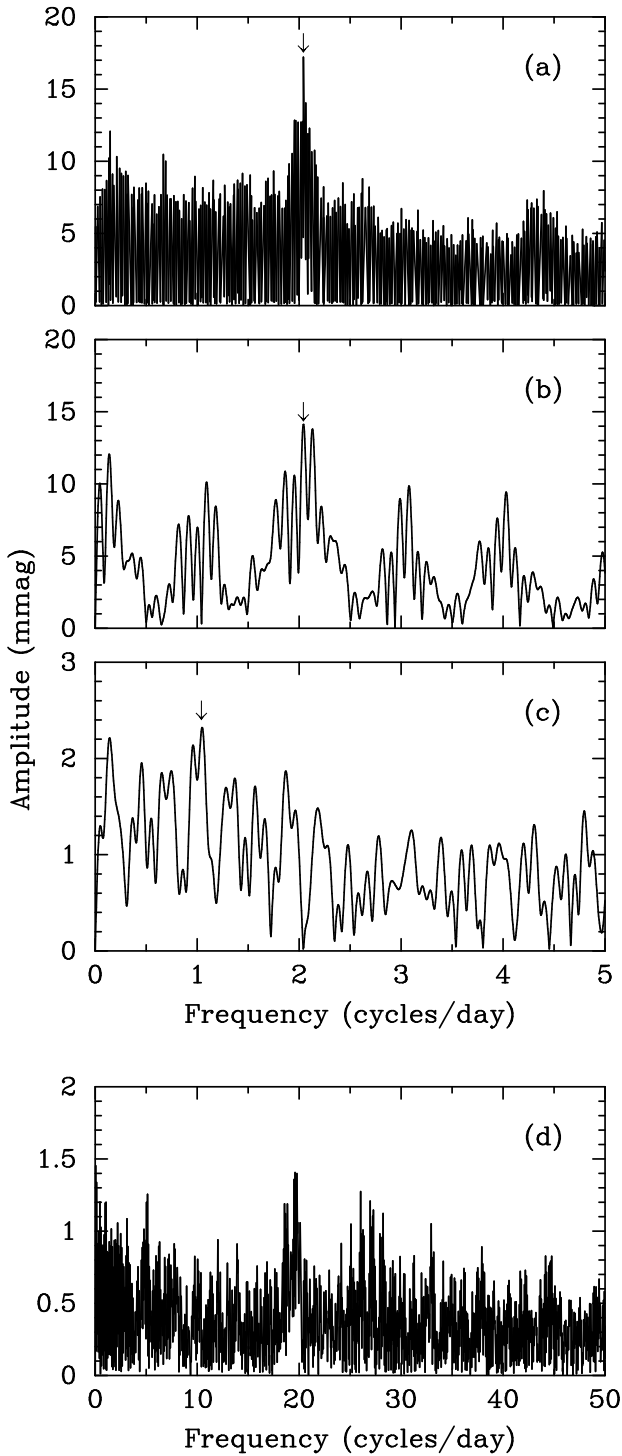}
\caption[]{(a): Amplitude spectrum of the {\it Hipparcos} photometry of HD
81421; the frequency in these data is indicated with an arrow. (b):
Amplitude spectrum of our ground-based V filter observations; the {\it
Hipparcos} period is confirmed. (c): Amplitude spectrum of our
measurements after prewhitening of the {\it Hipparcos} frequency. Residual
variation near 1 c/d is suspected. (d): Amplitude spectrum of our
measurements after prewhitening of the {\it Hipparcos} frequency and its
subharmonic. A wider range is chosen to show the suspected presence of
$\delta$ Scuti pulsations.}
\end{figure}

The frequency analysis of our data corroborated the {\it Hipparcos} period (cf.
Fig.~2 (b)), and the amplitudes in our B (19.0 $\pm$ 0.4) and V (14.9
$\pm$ 0.3) filter data suggest that this variability is due to pulsation.
However, after prewhitening this variation from the data, some residual
amplitude near 1.0 c/d, half the dominant frequency, remains. Such a
variation can of course also be caused by problem with the nightly
zeropoints or by imperfect extinction corrections etc. However, an
examination of the SAAO data only (which show excellent long-term
stability as judged from the measurement of the constant comparison
stars) suggests that this 1.0 c/d variation could be intrinsic to HD
81421. It could therefore be a subharmonic of the 11.75-h period,
indicative of ellipsoidal variation, but our data are insufficient to
confirm or reject this idea.

A search for $\delta$ Scuti pulsations of this star proved to be difficult
as well. Although a short-period variation with a time scale of about one
hour seemed to be superposed on the slow variability of the star, and
although the rms scatter of the nightly light curves of HD 81421 after
removing the long-term trends is higher than that of the difference of the
comparison stars, we cannot prove its presence. An amplitude spectrum of
all our data after prewhitening the {\it Hipparcos} frequency and its
suspected subharmonic (Fig.~2 (d)) shows evidence for low-amplitude
$\delta$ Scuti variability, but the signal-to-noise ratio is too small for
a definite detection.

On the basis of our data and those by the {\it Hipparcos} satellite we can
therefore still not pinpoint the physical nature of HD 81421. It could be
a $\gamma$ Doradus pulsator or an ellipsoidal variable, and it could
additionally be a $\delta$ Scuti star. More observations of the star are
needed. Photometric measurements with a suitable time distribution
and/or time series spectroscopy are required to understand this object.

\subsubsection{HD 85693} Although the {\it Hipparcos} photometry suggests
variability with a light range in excess of 0.1 mag, this star showed only
very small-amplitude slow variability during our observations. Therefore,
the error in the B/V colour amplitude ratio is too large to allow us to
pinpoint the cause of the light variations. We retain the star as a
$\gamma$ Doradus candidate and place an upper limit of 0.8 mmag on the
presence of $\delta$ Scuti pulsations in our light curves.

\subsubsection{HD 86371} We are unable to recover the {\it Hipparcos} period
suggested by Handler (1999) in our photometry, but the complicated light
curves and our measured B/V colour amplitude ratio (1.20 $\pm$ 0.03)
suggest pulsation as the reason for this star's variability. Our upper
limit for $\delta$ Scuti variability of this star is 0.8 mmag. The
multiperiodicity from the {\it Hipparcos} photometry, which is confirmed with a
residualgram analysis, and the colour amplitude ratio derived above lead
us to classify HD 86371 as a {\it bona fide} $\gamma$ Doradus star.

\subsubsection{HD 110606 and HD 113357} The light curves of both stars
appear quite complicated. Neither the {\it Hipparcos} data nor our
observations allowed the detection of a dominant period, which suggests
the presence of multiperiodic $\gamma$ Doradus pulsations. The B/V colour
amplitude ratios in our data are 1.26 $\pm$ 0.06 for HD 110606 and 1.34
$\pm$ 0.07 for HD 113357, further supporting this idea. No $\delta$ Scuti
variability is present within a limit of 1.0 mmag in either of the stars.
We cautiously classify HD 110606 and HD 113357 as prime $\gamma$ Doradus
candidates, but we think that further observations can easily prove they
are {\it bona fide} members of the group.

\subsubsection{HD 139095} We suggest that this is a {\it bona fide}
$\gamma$ Doradus star. The {\it Hipparcos} and the ground-based light
curves are multiperiodic, and the B/V colour amplitude ratio is 1.23 $\pm$
0.02, typical for pulsational variability of a late A/early F star. There
is no evidence for $\delta$ Scuti-type variability within a limit of 1.1
mmag in the amplitude spectrum.

\subsubsection{HD 167858} The {\it Hipparcos} data of the star imply it is
a multiperiodic $\gamma$ Doradus pulsator of quite high amplitude. This is
confirmed with our ground-based data, and the B/V colour amplitude ratio
(1.29 $\pm$ 0.06) implies pulsation as the cause for these light
variations. The search for $\delta$ Scuti pulsations of HD 167858 in two
nights yielded a null result, with no variation detected within 1.0 mmag
in the second night of our measurements, which was of much better quality
than the first one. We classify HD 167858 as a {\it bona fide} $\gamma$
Doradus star.

\subsubsection{BD+8 3658} In his search for multiperiodic variability
among {\it Hipparcos} variables, Koen (2001) found one frequency typical
for $\gamma$ Doradus variations (1.00064 cycles/day) and a second one
suggesting $\delta$ Scuti variability of BD+8 3658. This star was
therefore of considerable interest for us. Weak $\delta$ Scuti modulations
with a peak-to-peak amplitude of about 0.01 mag at a time scale of 2.5 h
($Q=0.035$\,d) and evidence for multiperiodicity are indeed present in our
data. However, we cannot identify the cause of the slow variability with
certainty. Our ground-based measurements show almost no slow variability,
a combined result of the {\it Hipparcos} period being very close to 1~d
and of insufficient phase sampling. Therefore we cannot calculate
meaningful colour amplitude ratios. However, the phase diagram of the {\it
Hipparcos} photometry relative to the long period (Fig. 3) is not typical
for pulsational variability. More observations of this star are needed,
including standard Str\"omgren photometry and spectroscopy.

\begin{figure}
\includegraphics[width=102mm,viewport=5 00 330 169]{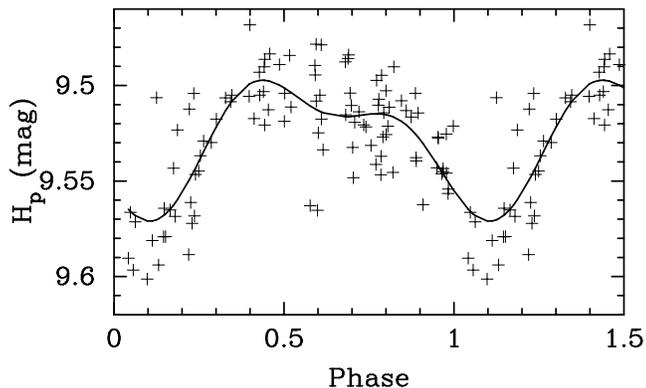}
\caption[]{The phase diagram of the {\it Hipparcos} photometry of BD+8 3658
relative to a frequency of 1.00064 cycles/day. The solid curve is the
result of a two-harmonic fit to these data. The light curve shape is
not indicative of pulsational variability.}
\end{figure}

\subsubsection{HD 173794} Handler (1999) could not decide whether this
star shows variations close to 1.5 days or of one or two hours because of
the aliasing problem in the {\it Hipparcos} data mentioned in the
Introduction. In fact, both types of variation are present, as already
seen from our first ground-based light curve (Fig. 4 (a)). There are clear
multiperiodic $\delta$ Scuti pulsations superposed on slow variations
which have very similar amplitude in the B and V bands (cf. Fig. 4 (b,
c)). This raises the suspicion that HD 173794 is an ellipsoidal variable
with a $\delta$ Scuti component. We therefore re-analysed the {\it
Hipparcos} photometry of that star with the residualgram method; the
result is shown in Fig. 4 (d). The dominant peak in this plot is at 0.3078
c/d, exactly 1/2 the frequency found by Fourier analysis (Handler 1999).
The phase diagram of the {\it Hipparcos} photometry relative to this
frequency is typical for an ellipsoidal variable, and the corresponding
frequency solution explains all the slow variability in these data.

\begin{figure}
\includegraphics[width=102mm,viewport=5 00 330 587]{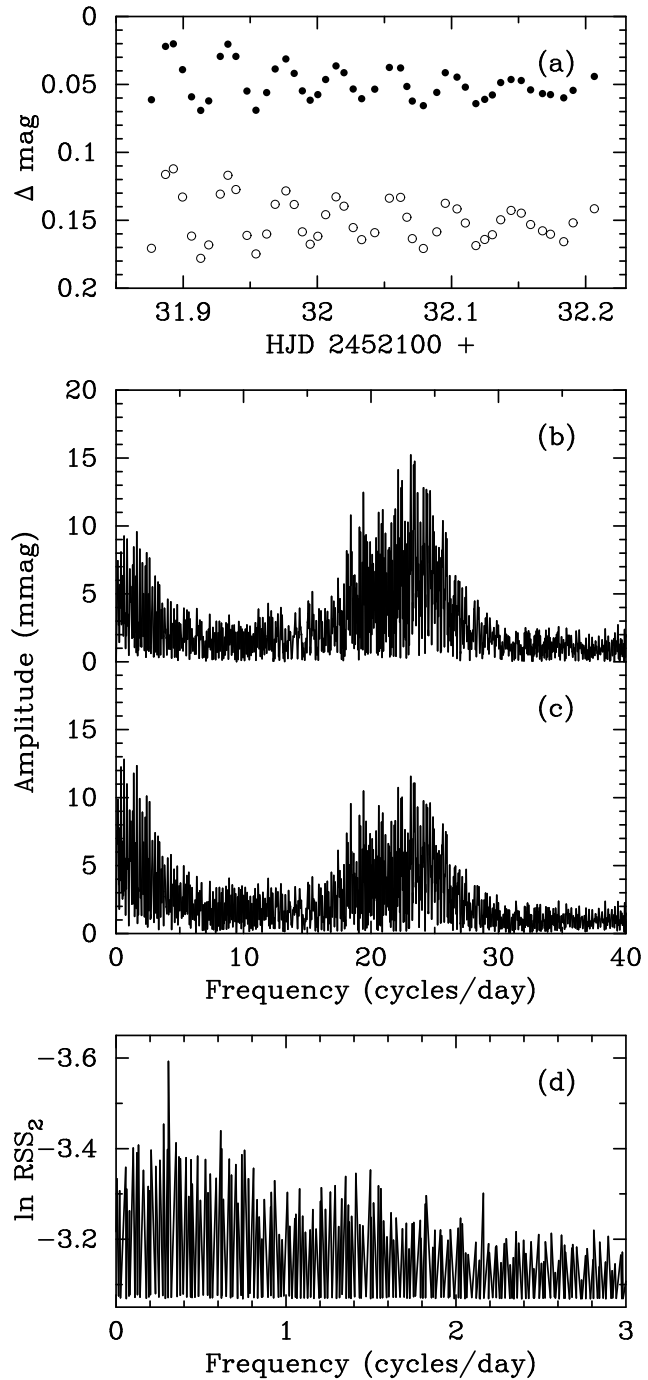}
\caption[]{(a): our first photometric measurements of HD 173794, showing
both multiperiodic $\delta$ Scuti pulsation and a slow drop in mean
magnitude (about 0.02 mag during this run). Filled circles are V filter
data, open circles the B filter measurements. (b): amplitude spectrum of
all our ground-based B filter observations; (c): amplitude spectrum of all
our V filter measurements. Note that the $\delta$ Scuti pulsations have
higher amplitude in the B filter, whereas the slow variability has
slightly higher amplitude in V. (d): residual sum of squares spectrum of a
2-harmonic fit (RSS$_2$) to the {\it Hipparcos} photometry of HD 173794.}
\end{figure}

These two frequencies also explain the slow variability in our data. We
have therefore fitted them to our measurements, adopting the {\it
Hipparcos} frequencies as definite and performed a frequency analysis of
the residuals to examine the $\delta$ Scuti variability in more detail.
The short-period pulsations turned out to be quite complicated, at least 5
frequencies in the range of 19 - 26 cycles/day appear to be excited. We
refrain from quoting exact values because of the aliasing problem. In any
case, the short-period pulsations are quite interesting. All evidence, the
{\it Hipparcos} parallax of the star, its spectral classification (A3 {\sc
iii-iv}, Houk 1978) and the long period of the ellipsoidal variation,
suggests it is rather evolved and thus pulsates in quite high radial
overtones ($Q \approx 0.006, k \approx 10$).

\subsubsection{HD 181998} We suggest that this star is a {\it bona fide}
$\gamma$ Doradus star. Our two longer nights of observations show slow
variability which is not singly-periodic, consistent with the complicated
amplitude spectrum of its {\it Hipparcos} photometry. These results
combined with the colour amplitude ratios ($B/V=1.22 \pm$ 0.03, $V/I_{\rm
c}= 1.68 \pm$ 0.03) in our data show that the slow variations of HD 181998
are due to pulsation. An upper limit of 1.1 mmag on the presence of
$\delta$ Scuti pulsations is placed.

\subsubsection{HD 188032} The analysis of the {\it Hipparcos} data of this
star (Handler 1999) left some doubt about it being a good $\gamma$
Doradus candidate or a more rapid variable. In our ground-based
observations, only slow variations with an amplitude below 0.02 mag are
detected. From these we infer a B/V colour amplitude ratio (1.25 $\pm$
0.1), suggestive of pulsation. Short-period variability in the $\delta$
Scuti domain remains undetected within 0.9 mmag. We classify this star as
a $\gamma$ Doradus candidate.

\subsubsection{HD 189631} This star was found to be slowly variable within
individual nights, but we are unable to determine the time scale of the
light variations. The colour amplitude ratios ($B/V = 1.36 \pm$ 0.04,
$V/I_{\rm c}= 1.82 \pm$ 0.12) imply pulsation as the cause of this
variability which reached a peak-to-peak amplitude in excess of 0.05 mag
from night to night. We conservatively classify HD 189631 as a $\gamma$
Doradus candidate and note the absence of $\delta$ Scuti pulsation
within an upper limit of 0.9\,mmag.

\subsubsection{HD 198528} We found slow variations with amplitudes of
several hundredths of a magnitude in our four longest observing runs.
However, only very small colour variability ($B/V$ amplitude ratio $<$
1.04) was noted as well, raising the suspicion that the star is an
ellipsoidal variable. A combined analysis of the {\it Hipparcos}
photometry (with the decisive clue coming from residualgram analysis, see
Fig. 5) and our new observations confirmed this interpretation. We found
that a double-wave light curve (inconsistent with W UMa-type variability)
corresponding to an orbital period of 0.807 d satisfactorily explains all
the data. The period given by Handler (1999) based on {\it Hipparcos}
photometry only is therefore incorrect, probably due to the small number
of observations (49 measurements distributed over 3 years). No $\delta$
Scuti variation is detected within a limit of 1.5 mmag in the amplitude
spectrum, although we note that a very weak 45-minute variation seemed to
be present in both filters in our best data.

\begin{figure}
\includegraphics[width=102mm,viewport=5 00 330 319]{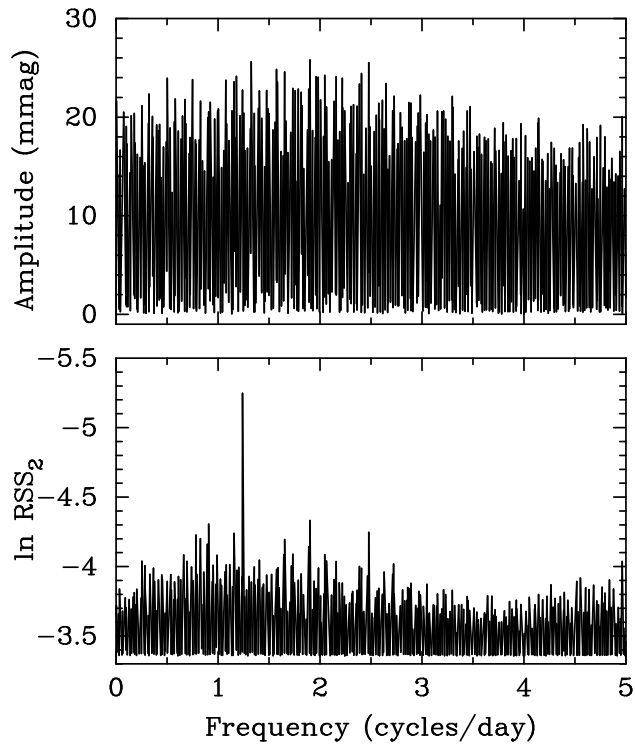}
\caption[]{Upper panel: amplitude spectrum of the {\it Hipparcos} photometry of
HD 198528. No periodicity can be found. Lower panel: residual sum of
squares spectrum of a 2-harmonic fit (RSS$_2$) of the same data. The
correct frequency of light variation is now quite clear.}
\end{figure}

\subsubsection{HD 201985} This star remains unsolved. Although the {\it
Hipparcos} photometry implies a range of variability of a few hundredths
of a magnitude, little variation was seen during our observations (a few
millimagnitudes at best), with one exception: during the first run from
SAAO, the star was 0.15 mag fainter as in the other nights. An
instrumental problem or misidentification on the sky is ruled out, as the
star's mean $(B-V)$ colour on that night was the same as in the other four
within fractions of a millimagnitude, and as the relative zeropoints of
the other three stars in the ensemble were consistent with the other
nights. We note that little colour variability seems to accompany the
magnitude changes, and we can place an upper limit of 2.0 mmag to the
presence of $\delta$ Scuti pulsations. HD 201985 could be an eclipsing
binary.

\subsubsection{HD 207223} This is a singly-periodic {\it bona fide}
$\gamma$ Doradus star (Aerts \& Kaye 2001) located close to HD
207651 in the sky. We tested HD 207223 for $\delta$ Scuti variability in a
single night. None was detected within a limit of 1.5 mmag in the
amplitude spectrum (cf. Kaye et al. 1999b).

\subsubsection{HD 207651} The {\it Hipparcos} photometry seemed to indicate two
frequencies of 1.4 and 6 c/d (Handler 1999), whereas our runs show a
dominant $\delta$ Scuti variation with a time scale of 1.5 - 2 hr with
superposed long-term modulations. The associated colour amplitudes
corroborate the $\delta$ Scuti interpretation, but are inconclusive with
respect to the slower variability. Estimating the absolute magnitude of the
star from its {\it Hipparcos} parallax results in $M_{\rm v} = -0.4 \pm 0.5$,
but Str\"omgren photometric calibrations (Crawford 1979) yield $M_{\rm v}
= +0.8 \pm 0.3$. This may be an indication of binarity. In any case, we
are unable to pinpoint the nature of the star on the basis of our data
alone; more observations are needed. The $Q$ values of the $\delta$ Scuti
pulsations are 0.007 and 0.015 d for the two absolute magnitude values,
respectively.

\subsubsection{HD 209295} This is our only detection of both $\delta$
Scuti and $\gamma$ Doradus pulsations in the same star. It was already
clear from the analysis of the {\it Hipparcos} photometry (Handler 1999)
that multiperiodic $\gamma$ Doradus variability is present in the star's
light curves, and multiperiodic $\delta$ Scuti pulsations were clearly
detected in our first nights of observation as well (Fig.~6). HD 209295
was studied in great detail by Handler et al. (2002), and we refer to this
paper for more information. For the present purposes, we only note that
these authors present evidence that the $\gamma$ Doradus pulsations of HD
209295 are tidally excited. The star should therefore not be considered a
normal $\gamma$ Doradus star.

\begin{figure}
\includegraphics[width=102mm,viewport=5 00 330 170]{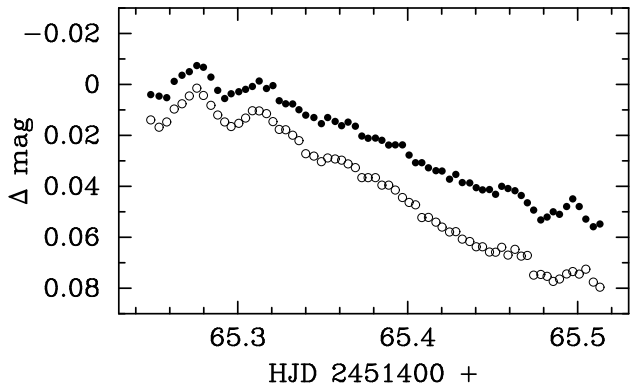}
\caption[]{An example light curve of HD 209295. Filled circles are V
filter data, open circles the B filter measurements. Multiperiodic
$\delta$ Scuti pulsations are superposed on a gradual decline in
brightness. Note the higher amplitude of the slow variations in the B
filter data, confirming their pulsational origin.}
\end{figure}

\subsubsection{HD 211699} This prime $\gamma$ Doradus candidate (Handler
1999) showed very little variability during our individual nights of
observation which may be due to the dominant period of light variation
being close to 1 day. However, we were able to detect changes in the mean
magnitude from night to night, whose B/V colour amplitude ratio (1.29
$\pm$ 0.09) suggests that pulsations are causing them. We therefore retain
this star as a $\gamma$ Doradus candidate and we also note the absence of
$\delta$ Scuti pulsations within an upper limit of 0.8 mmag.

\subsubsection{HD 221866} Classified as a prime $\gamma$ Doradus candidate
by Handler (1999), this star showed clear evidence of slow variability.
The {\it Hipparcos} period fits our data and results in a B/V colour amplitude
ratio of 1.18 $\pm$ 0.02, consistent with pulsation. $\delta$ Scuti-type
variability remains undetected within a limit of 0.9 mmag in the amplitude
spectrum. We conservatively retain this star as a $\gamma$ Doradus
candidate because of the rather low colour amplitude ratio.

\subsubsection{Variable comparison stars}

Out of the 40 comparison stars used in this study, two turned out to be
variable as well. The first, HD 86301, was a comparison star in the HD
85693/HD 86371 group. It is located near the hot luminous border of the
$\delta$ Scuti instability strip and was found to be a very low-amplitude
variable. The light curve appears multiperiodic with a time scale of 3.8
-- 6 h and a total amplitude of about 0.01 mag. Our period estimate yields
a range of 0.035 d $< Q <$ 0.054 d. We tentatively classify HD 86301 as a
new $\delta$ Scuti star.

The second variable comparison star was HD 183452, chosen for the HD
181998 group. During our first short run on this group, it already showed
conspicuous magnitude changes. After realising the star was classified as
variable from measurements with the {\it Tycho} satellite (ESA 1997), we
added another comparison star into the sequence, but continued to monitor
HD 183452 as well. The total amplitude of its light variations is in
excess of 0.1 mag, but there is very little colour variation. We thus
suspect that HD 183452 is an ellipsoidal variable. A double-wave light
curve assuming a preliminary orbital period of 0.692\,d fits our data very
well.

\subsection{Summary of survey results}

We performed photometric monitoring of a total of 26 stars which seemed to
be related to the $\gamma$ Doradus phenomenon. One of them, HD 209295,
turned out to be both a $\gamma$ Doradus and a $\delta$ Scuti pulsator,
but this object is peculiar, as its g-mode pulsations appear strongly
coupled to its binary orbit (Handler et al. 2002).

We believe that six of our targets are new $\gamma$ Doradus pulsators, and
nine more stars are probable $\gamma$ Doradus stars. We discovered three
ellipsoidal variables and one $\delta$ Scuti star. The classifications of
our target stars are summarized in Table\,2.

\begin{table*}
\caption[]{Programme star classifications. Objects indicated with an
asterisk are $\delta$ Sct stars in addition to their main type of
variability, whereas stars indicated with two asterisks have suspected
additional $\delta$ Sct variations.}
\begin{center}
\begin{tabular}{cccccc}
\hline
$\gamma$ Dor/$\delta$ Sct & {\it bona fide} & $\gamma$ Dor &
ellipsoidal & $\delta$ Sct stars & unsolved \\
``hybrid'' & $\gamma$ Dor stars & candidates & variables & & variables \\
\hline
HD 209295 & HD 12901 & HD 10167 & HD 14147$^{\ast\ast}$ & HD 27093 & HD
81421$^{\ast\ast}$ \\
 & HD 65526 & HD 40745 & HD 173794$^{\ast}$ & & HD 86593 \\
 & HD 86371 & HD 41448 & HD 198528$^{\ast\ast}$ & & BD+8 3658$^{\ast}$ \\
 & HD 139095 & HD 110606 & & & HD 201985 \\
 & HD 167858 & HD 113357 & & & HD 207651$^{\ast}$ \\
 & HD 181998 & HD 188032 & & & \\
 & HD 207223 & HD 189631 & & & \\
 &  & HD 211699 & & & \\
 &  & HD 221866 & & & \\
\hline
\end{tabular}
\end{center}
\end{table*}

The detection limits we achieved for $\delta$ Scuti pulsations compare
well with those of other ground-based variability surveys (starting with
Breger 1969), to which our results are to be compared. Of course, the
presence of very low-amplitude $\delta$ Scuti pulsations in some of our
targets cannot be ruled out. Such variability has been detected down to
limits of 0.4 mmag with the help of extensive multisite campaigns (Handler
et al. 2000), and space missions are expected to decrease these limits
considerably.

\section{The hot border of the $\gamma$ Doradus phenomenon}

As already mentioned in the Introduction, it is not quite clear up to what
effective temperatures $\gamma$ Doradus pulsations can be excited. With
our new results added to those available in the literature, we can take a
closer look at this problem. We show all the presently known $\gamma$
Doradus stars and suspects (taken from the on-line catalogue by Handler \&
Kaye 2001) in a colour-magnitude diagram in Fig.~7.

\begin{figure*}
\includegraphics[width=103mm,viewport=60 00 485 360]{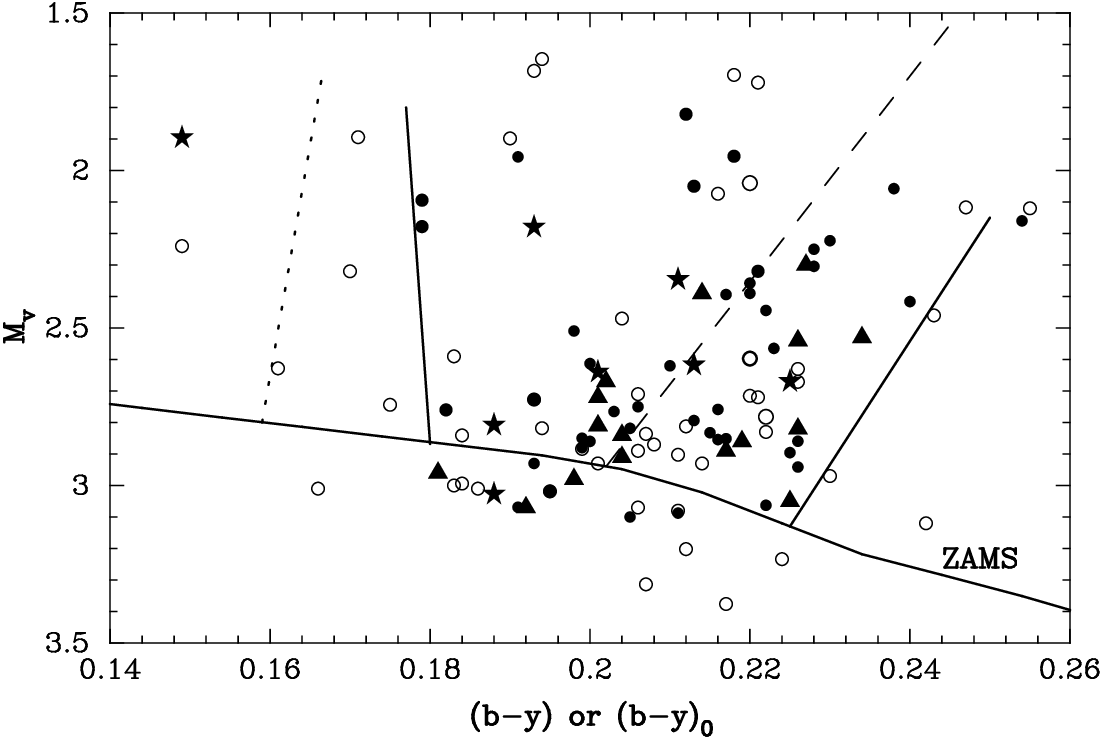}
\caption[]{The domain of the $\gamma$ Doradus pulsators in the
colour-magnitude diagram. Star symbols denote {\it bona fide} $\gamma$
Doradus stars observed by us and filled triangles are {\it bona fide}
$\gamma$ Doradus stars from the literature. The filled circles are prime
$\gamma$ Doradus candidates and the open circles are other $\gamma$
Doradus candidates (including our unsolved variables). Ellipsoidal
variables have been omitted. The approximately horizontal line is the
Zero-Age Main Sequence (Crawford 1975, 1979), and the dotted line is the
blue boundary of the $\gamma$ Doradus region derived by Handler (1999),
which is superseded by this work. The lines almost normal to the ZAMS are
the presently known boundaries of the $\gamma$ Doradus region. The dashed
line almost normal to the ZAMS represents the red edge of the $\delta$
Scuti instability strip (Rodriguez \& Breger 2001). The $\gamma$ Doradus
star outlying far to the blue is HD 209295.}
\end{figure*}

Even a brief glance at Fig.~7 suggests that the blue edge of the $\gamma$
Doradus instability region as outlined by Handler (1999) requires
revision. (We again exclude HD 209295, the hottest {\it bona fide}
$\gamma$ Doradus star from the discussion because of its close-binary
nature.) The three hottest stars referred to by Handler (1999) were HD
152569 (Kaye 1998), 57 Tau (Papar\'o et al. 2000) and M34 UVa 144
(Krisciunas \& Patten 1999). HD 152569 was reclassified as a $\delta$
Scuti star (Kaye, Henry \& Rodriguez 2000); thus it is not plotted in
Fig.~7.

57 Tau is also a $\delta$ Scuti star, but it exhibits slow, apparent
multiperiodic, variability with an amplitude of a few mmag (Papar\'o et
al. 2000) in addition; the period of the highest-amplitude slow variation
is 1.246 days. However, Kaye (1999) showed that 57 Tau is a spectroscopic
binary with an orbital period of 2.486 days, almost exactly twice the
value from the photometry. As there seems to be very little colour
variation in the slow variations noted by Papar\'o et al. (2000), we think
that 57 Tau is in fact an ellipsoidal variable. Therefore we do not any
longer consider the star a good $\gamma$ Doradus candidate.

The star UVa 144 in the open cluster M34 is a multiperiodic slow variable.
However, the two periods found by Krisciunas \& Patten (1999), 1.52 and
1.28 cycles/day, could be harmonics of each other allowing for some
aliasing ambiguities. In any case, the ($b-y$) colour of this star adopted
by Handler (1999) is erroneous, as he was unaware of the measurements by
Canterna, Perry \& Crawford (1979) showing the star to be redder than
previously thought.

Consequently, a revision of the blue boundary of the $\gamma$ Doradus
region in the colour-magnitude diagram is in order. The new blue edge is
already included in Fig.~7. Compared to the original one by Handler
(1999), it is about 150 K cooler on the ZAMS and about 75 K cooler one
magnitude above it. There are now four stars which define it, the {\it
bona fide} variable HD 218396 (Zerbi et al. 1999), as well as HD 41448
(Eyer 1998, Handler 1999 and this work), HD 211699 and HD 221866 (Handler
1999 and this work). As the evidence for the $\gamma$ Doradus nature of
all these stars is quite good, we think that another shift of this blue
edge towards redder colours is no longer possible.

\section{The incidence of $\delta$ Scuti pulsations amongst $\gamma$
Doradus stars}

With the exception of the unusual variable HD 209295, we have observed
five {\it bona fide} $\gamma$ Doradus stars located inside the $\delta$
Scuti instability strip (HD 12901, HD 65526, HD 86371, HD 167858 and HD
181998) adopting the red edge for $\delta$ Scuti pulsations by Rodriguez
\& Breger (2001)\footnote{This revised red edge only became available near
the end of this work and could therefore not be adopted for target
selection (Sect. 2).}. Another star, HD 139095, falls on this red edge.

We did not find $\delta$ Scuti pulsations in any of these stars. Does this
mean that, except under special circumstances such as the close binarity
of HD 209295, $\gamma$ Doradus stars cannot be $\delta$ Scuti pulsators at
the same time?

As mentioned in the Introduction, the average incidence of photometrically
detectable $\delta$ Scuti pulsations of chemically normal stars located
inside the lower instability strip is 1/3. However, the $\gamma$ Doradus
star domain only overlaps with a region close to the cool edge of the
$\delta$ Scuti instability strip. Hence it is quite possible that the
incidence of $\delta$ Scuti pulsations in that overlap region is generally
smaller than, for instance, in the middle of the $\delta$ Scuti strip.

To examine this idea, we assume that all stars tested for $\delta$ Scuti
pulsations so far are uniformly distributed over $(b-y)$. We selected all
249 $\delta$ Scuti stars from the recent catalogue of Rodriguez,
L\'opez-Gonz\'alez \& L\'opez de Coca (2000), for which uvby$\beta$
photometry is available and determined their distance from the red edge of
the $\delta$ Scuti instability strip, as defined by Rodriguez \& Breger
(2001). We then determined the number of stars in strips 0.01 mag wide in
$(b-y)$, which are parallel to the red edge, and we show the number of
{\it bona fide} $\gamma$ Doradus stars we observed in those bins (Fig.~8).

\begin{figure}
\includegraphics[width=100mm,viewport=-05 00 335 180]{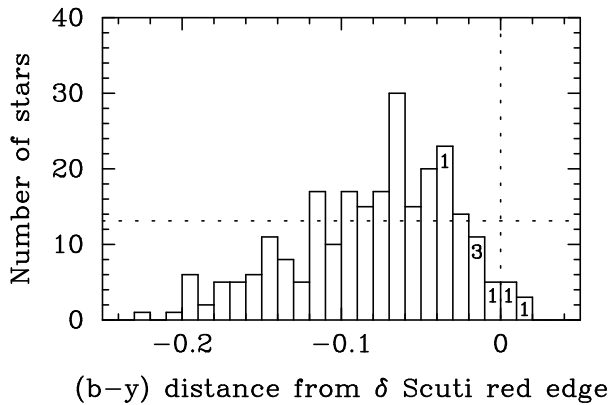}

\caption[]{The number of $\delta$ Scuti stars as a function of ($b-y$)
distance from the red edge of the lower instability strip (vertical dotted
line). In this diagram, the blue edge is located at $\Delta (b-y)=-0.19$
on the zero-age main sequence and at $\Delta (b-y)=-0.23$ at the luminous
end of the $\delta$ Scuti strip. The horizontal dotted line is the average
number of stars per bin for $-0.19 < \Delta (b-y) < 0$. The numbers inside
some of the bins are the number of {\it bona fide} $\gamma$ Doradus stars
tested for $\delta$ Scuti variability corresponding to their $\Delta
(b-y)$.}
\end{figure}

As suspected, the number of $\delta$ Scuti stars decreases towards the red
edge; stars near this red edge have an incidence of $\delta$ Scuti
pulsation which is lower than average. We need to take this into account
if we want to examine the significance of the absence of $\delta$ Scuti
pulsation in the {\it bona fide} $\gamma$ Doradus stars we observed.

Thus, the probability that we do not find $\delta$ Scuti pulsations in any
of our observed {\it bona fide} $\gamma$ Doradus stars inside the $\delta$
Scuti strip, results in 18\%. This suggests that $\gamma$ Doradus stars
are less likely to be $\delta$ Scuti pulsators than non-$\gamma$ Doradus
stars in the same domain of the colour-magnitude diagram. Obviously, more
stars need to be observed to strengthen this conclusion.

\section{Pulsation periods and pulsation constants}

The longest pulsation periods of $\delta$ Scuti stars listed in the
catalogue of Rodriguez et al. (2000) are around 6.5 h; the possible
$\delta$ Scuti stars AC And (Fernie 1994) and V823 Cas (Antipin 1997) even
have periods of 17 and 16 h, respectively. The shortest pulsation periods
of $\gamma$ Doradus stars found so far were around 7.5 h (Handler 1999,
Henry et al. 2001). This may lead to the suspicion that there might be an
overlap in the pulsational behaviour of those two classes of pulsator,
making a distinction hard for certain stars.

However, as the $\gamma$ Doradus stars are so far only found on the main
sequence, and as the long-period $\delta$ Scuti stars all seem to be
evolved, this overlap is not a physical one. To support this suggestion,
we compared the distribution of the pulsation periods of 636 $\delta$
Scuti stars in the catalogue of Rodriguez et al. (2000) to that of the
{\it bona fide} stars (except HD 209295) listed by Handler \& Kaye (2001).
We show it in the upper panel of Fig.~9. As implied by the previous
discussion, there is almost an overlap between the two groups.

On the other hand, if one uses the pulsation constant $Q$ (calculated with
Eq. 1 and the method described in Sect. 3.1) to discriminate between
$\delta$ Scuti and $\gamma$ Doradus stars, the ambiguity is removed (lower
panel of Fig. 9, 262 $\delta$ Scuti stars from Rodriguez et al. (2000) for
which pulsation constants could be calculated). There is a clear gap
between the two groups. Thus we confirm that $\delta$ Scuti and $\gamma$
Doradus stars can be separated via their pulsation constants; all {\it
bona fide} $\gamma$ Doradus stars known to date have $Q>0.23$~d.

\begin{figure}
\includegraphics[width=95mm,viewport=-04 00 344 405]{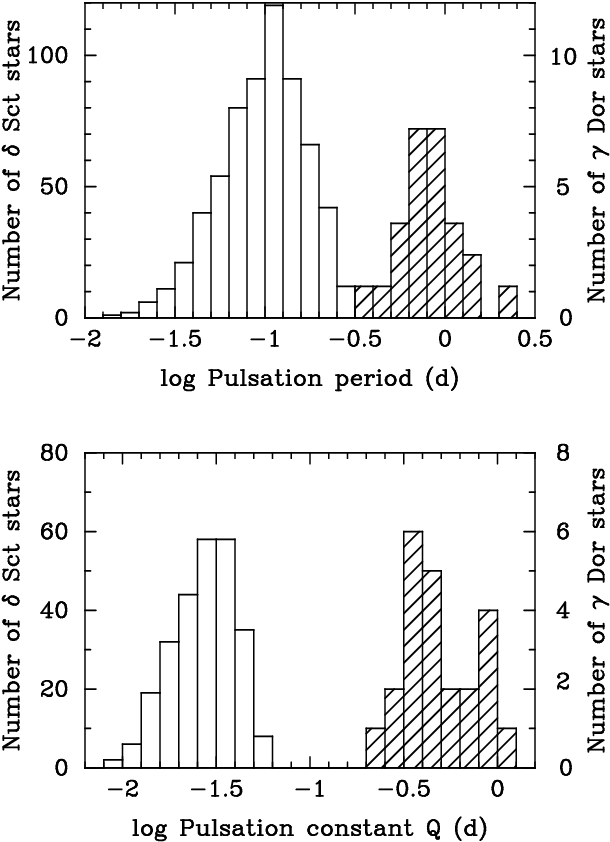}
\caption[]{Upper panel: the number of $\delta$ Scuti (open histogram bars)  
and $\gamma$ Dor (hatched histogram bars) stars within certain ranges of
pulsation period. The two groups of pulsator almost overlap in this
diagram. Lower panel: the distribution of the pulsation constants of
$\delta$ Scuti (open bars) and $\gamma$ Dor (hatched bars) stars. There is
a clear separation between the two.}
\end{figure}

\section{Conclusions}

We performed a photometric search for $\delta$ Scuti pulsations among
candidate $\gamma$ Doradus stars to examine the interrelations between
these two groups of pulsator. We observed altogether 26 stars, of which
65\% turned out to be {\it bona fide} $\gamma$ Doradus stars or excellent
candidates. However, none of these objects exhibited $\delta$ Scuti
pulsations with the exception of the close binary HD 209295 (Handler et
al. 2002), which is therefore anomalous.

Our results indicate that $\gamma$ Doradus stars are less likely to
be $\delta$ Scuti pulsators than non-$\gamma$ Doradus stars in the same
temperature range. This conclusion is not yet definite because of the
small number of {\it bona fide} $\gamma$ Doradus stars inside the $\delta$
Scuti domain investigated so far. This situation can be improved by
performing a similar project in the Northern Hemisphere and by more
extensive observations of the $\gamma$ Doradus candidates we already
examined, in order to prove their $\gamma$ Doradus nature.

We were also able to locate the blue edge of the $\gamma$ Doradus domain
in the colour-magnitude diagram more accurately. Finally, we showed that
the $\delta$ Scuti stars and the $\gamma$ Doradus pulsators are clearly
separated by the values of their pulsation constants $Q$; the known
$\gamma$ Doradus stars all have $Q>0.23$~d. 

Although we are still at the beginning of understanding the whole extent
of the $\gamma$ Doradus phenomenon, we think we now have the basic data
for quantitative comparisons between observations and model calculations.
Acceptable models for $\gamma$ Doradus pulsators must be able to reproduce
the observed constraints on pulsational instability in temperature,
luminosity, metallicity, and period.

\section*{ACKNOWLEDGEMENTS}

GH thanks Eloy Rodriguez for making an ASCII version of his $\delta$ Scuti
star catalogue available and Chris Koen for his support of this project
and for carefully proofreading a draft version of this paper. Michel
Breger, Laurent Eyer and Tony Kaye also commented on a draft version of 
this work. We appreciate the constructive comments of the referee, Joyce
Guzik.

\end{document}